# A Comparative Study for Block Chain Applications in The Manet

Sangheethaa S

College of Information Technology, University of Fujairah, Fujairah, UAE

## Abstract

*MANET- Mobile Ad-hoc Networks are famous for their infrastructure-less arrangement for communication. In this network, nodes are self-organized and can act as router. They are battery operated and self-organizing. Block chain is a new concept from 2008 and researchers are trying the possible application of Block chain in many sectors including MANETs. This paper surveys the existing researches done in applying block chain in a MANET environment. Block chain is mainly used in MANETs for improving security while routing packets from one node to another. Some researchers have proposed trust models using block chain. This paper reviews some of the existing approaches where block chain is used in MANETs for routing the packets, creating trust models, and dealing with network partitioning problem and scalability problem. This paper acts as a review paper to study on block chain applications in MANET.*

## 1. Introduction

Mobile Ad hoc Networks (MANETs) [1] are a type of wireless ad hoc networks that enable mobile devices to communicate with each other without relying on a fixed infrastructure or centralized control. In MANETs, each device acts as a node that can send, receive, and forward data packets to other nodes in the network.

The nodes in the MANET are laptops, smartphones, or tablets, equipped with wireless interface for communication. These devices can form a network by establishing direct communication links with nearby nodes, and by relaying data packets to nodes that are out of range.

MANETs are particularly useful in situations where a fixed infrastructure is unavailable or unreliable, such as in emergency and disaster response scenarios, military operations, and remote areas where there is no existing communication infrastructure. They are also useful in environments where devices need to communicate with each other without relying on a centralized control, such as in vehicular networks and Internet of Things (IoT) applications.

However, MANETs face several challenges, such as limited bandwidth, battery life, and security threats. The limited bandwidth and battery life of mobile devices can affect the performance and reliability of the network, while the lack of a centralized control and the dynamic nature of the network make it vulnerable to security threats such as eavesdropping, Denial of Service Attack and packet drops by malicious nodes.

Various researches have been done to improve the MANET environment by means of Routing, congestion control and security mechanisms. This paper gives a review of existing researches which finds their application of Blockchain in MANET environment. The paper is organized as follows. The next 2 sections give the introduction to MANET and Blockchain . Section 4 gives the already existing review or comparative study papers in this area. Section 5 to 9 gives review of the research papers as on date which deals with Blockchain and MANET finally the conclusion section. Table 1 gives the summary of research works discussed in this paper.

Table 1 Comparison table

| Paper | Use of blockchain for | Remarks |
| --- | --- | --- |
| [2] | Trust control | Use of PBFT, is introduced. |
| [3] | Trust management | Introduced concept of new consensus model called as Delegated Proof of Trust mechanism |
| [4] | Routing | Introduced approach to calculate reputation of a node and sharing it with others over AODV Protocol. |
| [5] | Network partitioning problem | Uses the DAG (Directed Acyclic Graph)structure in a permissioned Blockchain. |
| [6] | Routing | Enhances OLSR protocol to use reputation based on blockchain ledger concept. |
| [7] | Growth problem | By using a special genesis block to control the growth to reduce the disk space usage. |

## 2. BLOCKCHAIN BACKGROUND

Blockchain is a distributed ledger technology that enables secure, transparent, and tamper-resistant record-keeping of transactions or data. It was originally introduced in 2008 as a foundational technology for Bitcoin [2], a decentralized digital currency system, but now it is applied to various fields where the aim is to eliminate a third party. At its core, a blockchain is a digital ledger that records a series of transactions or data blocks in a chronological and immutable manner. Each block contains a hash of the previous block, forming a chain of blocks, hence the name "blockchain". Once a block is added to the chain, it cannot be modified or deleted without invalidating all subsequent blocks, making it resistant to tamper and fraud.

Blockchain technology [3] is based on a distributed network of nodes, where each node maintains a copy of the ledger and participates in the validation of new transactions or blocks. Transactions are validated by the nodes through a consensus mechanism, which ensures that all nodes agree on the validity of the transactions before they are added to the blockchain.

Figure 1 shows the principles of blockchain technology. The consensus mechanism used in public Blockchain are categorized into 3. Proof of work, Proof of stake, delegated proof of stake. Details of these mechanisms can be seen from many research work like [4]. The basis of blockchain is the complex cryptographic algorithms.

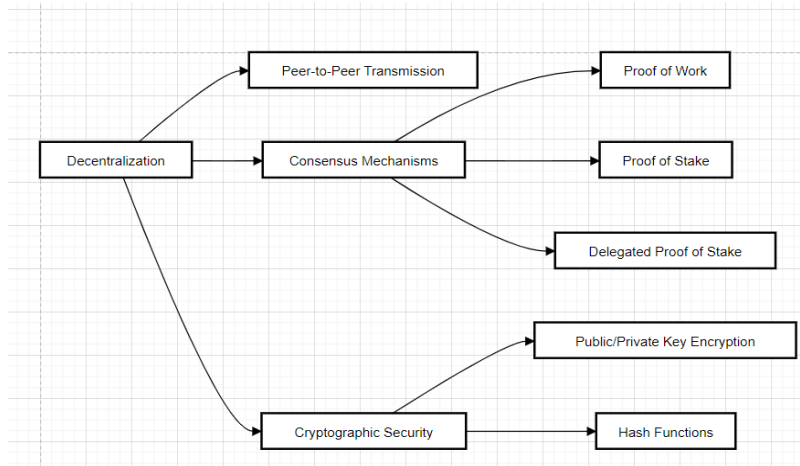

Figure 1 Principles of Blockchain (drawn using draw.io)

The decentralized and transparent nature of blockchain technology offers several advantages [5], including increased security, transparency, and efficiency. It eliminates the need for intermediaries or central authorities, reducing costs and improving trust between parties. It also enables new forms of peer-to-peer transactions, smart contracts and De-centralized Applications(DAAPs)

Blockchain has found its place in supply chain management, logistics management, Healthcare data privacy, educational genuineness verification etc.

## 3. LITERATURE REVIEW

There are researches done in application of Blockchain in MANETS, VANETS and also the new kind of network called FANETs. These researches are focusing on how blockchain technology can be applied for improving performance or improving security in Mobile Ad hoc Networks or Vehicular Networks or Internet of Things.

The authors of [6] have given a comprehensive survey of application of Blockchain in Vehicular networks. They have given a detailed review of blockchain based VANETs, security constraints to be considered, challenges and simulation tools that could be used in VANET environment to test Block chain.

In [7] the authors have done a systematic comparison of application of Blockchain in Vehicular networks. [8] gives a detailed comparison of approaches for incentive based data forwarding in MANETs and approaches to use blockchain for data forwarding.

Authors of [9] gives the research directions and guidelines for the authors to use blockchain technologies in IoTs, MANETS and VANETS.

The author of this paper reviews about applications of Blockchain in MANET environment specifically related to security, trust management and scalability issues.

## 4. USING BLOCKCHAIN TECHNOLOGY IN MANETS SECURITY

The paper [10] explores the use of blockchain technology for trust control among nodes of MANETs. The authors give explanation of MANET security challenges and Blockchain technology. They also talk about the limitations of Blockchain when applied in MANET. They discuss the disadvantages of using Proof of stake and Proof of work in a MANET environment. Proof of Work is more computing intensive, Proof of Stake is challenging in MANET environment because of the highly dynamic nature of MANET.

Some blockchain technologies like hyperledger are using PBFT- Practical Byzantine Fault tolerance. This approach requires 2 out of 3 nodes to agree on the agreement. In MANET scaling may be an issue due to constant breaking of links. So PBFT may not be effective. The authors also analyzed about applying blockchain based concepts for trust management, which is discussed in the next section.

## 5. BLOCKCHAIN- BASED LIGHTWEIGHT TRUST MANAGEMENT IN MOBILE AD-HOC NETWORKS

Blockchain-Based Lightweight Trust Management is a technique used in Mobile Ad-Hoc Networks (MANETs) [11] to establish trust between nodes in a decentralized and distributed manner. In MANETs, trust management is a crucial factor to ensure secure communication among the nodes. The traditional centralized trust management systems are not suitable for MANETs as they require a centralized authority to manage the trust, which is not practical in a decentralized network. Therefore, a decentralized approach is needed to manage trust in MANETS.

Blockchain-Based Lightweight Trust Management (BLTM) is a technique used in Mobile Ad-hoc Networks (MANETs) to improve the security and reliability of communication among nodes. BLTM combines the trust management technique with blockchain technology to establish a secure and transparent trust model among nodes.

BLTM protocol for MANETs, which consists of four phases as shown in Figure 2. They are trust evaluation, blockchain-based consensus, block generation, block maintenance phase.

In the trust evaluation phase, each node evaluates the trustworthiness of its neighboring nodes based on various parameters such as the packet forwarding rate, response time, and packet drop rate. The node then assigns a trust value to each neighboring node based on the evaluation.

In the blockchain-based consensus phase, the node shares its local blockchain with the neighboring nodes to reach a consensus on the trust values assigned to each node. The consensus mechanism used in BLTM is a lightweight consensus mechanism that reduces the computational overhead and ensures the consensus is reached in a timely manner. It is named as Delegated Proof of Trust mechanism (DPoT). It works with OLSR protocol. They have adopted the DCFM scheme to develop DPoT. DPoT uses the Validator and delegator nodes for achieving consensus.
In the block generation phase, to create a block in a blockchain system, it's necessary to determine what information will be included in the block and how it will be configured by the delegate node. Once transactions in the pool are collected into a block, the blockchain system creates a hash value using the SHA-256 algorithm, which directly comes from the transaction data, and appends it to the block. The previous block's hash is also included as data in the current block to link the blocks together, creating a chain. The block hash is designed to accept only a specific format, such as a hash signature that starts with 10 consecutive zeros.

In a secure Mobile Ad-Hoc Network (MANET), a blockchain can be used to ensure trust between nodes. Each block in the blockchain contains a set of transaction data and metadata, which includes a timestamp, transaction hash, delegate ID, and nonce. When a transaction is hashed, the transaction generator ID and related transaction values are included, along with the delegate ID. This helps to ensure that block transactions are trustworthy and cannot be repudiated by any of the participating nodes. The first block in the blockchain, known as the "genesis block," is created with an empty list of transactions when the network is initially formed.

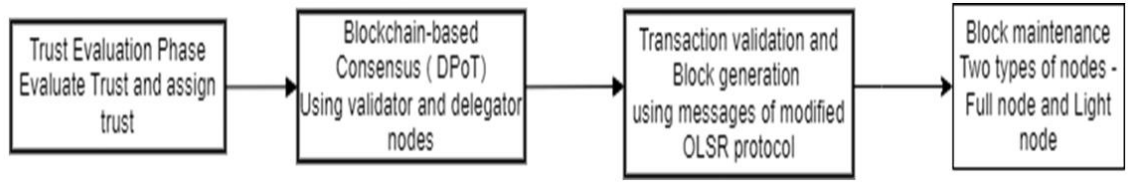

Figure 2. Phases of BLTM Protocol

## 6. REPUTATION BASED ROUTING IN MANETS USING BLOCKCHAIN

Maqsood Ahamed et al proposed a reputation-based routing protocol that uses blockchain technology in MANETs [12]. This protocol is implemented over AODV [14] protocol. This paper redefines the cost calculation of AODV protocol. They propose new link cost between the nodes using reputation score. The most reputed path is selected for routing to avoid malicious nodes in the route. Reputation is maintained in the blockchain and it is validated by part of the network nodes called as miners. A reputation value is assigned to nodes, based on their behavior in participating network operations. A miner, monitors the transaction done by the nodes in its vicinity and classifies them as good or bad. These transactions are combined together to form a block. They claim that the use of blockchain in routing has 2 advantages. 1. It provides immutable record of behavior of the nodes. 2. It acts as a gauge to validate credibility of the miners, by increasing difficulty level [12].

Anyhow, the paper does not talk more about the consensus mechanism followed for agreeing upon a block. The need for such immutable record in MANET environment is also not justified.

## 7. BLOCKGRAPH

Blockgraph [13] is a blockchain-based framework designed for Mobile Ad hoc Networks (MANETs). The goal of Blockgraph is to use DAG (Directed Acyclic Graph) based structure to avoid network partition problem. The Blockgraph framework, consists of three components: the Consensus system, the Block Management System and the Group management system[13]. The consensus system is avote based consensus mechanism that can be used in Blockgraph, as Block graph is a permissioned network. There are 2 sub systems in consensus system- 1. Leader Election 2. Log replication. Every network partition will get a new leader. All the nodes in the network are assumed to be trusted. The block is propagated to all peers by the leader node. The Block management system takes care of the local block structure including Creation, validation and ordering of blocks, recovery of missing blocks. The Group management system- runs in every node and detect if any change happens in the topology. Groups are reformed as a result of topology change. The main objective of this paper is to use Blockgraph to solve the issue of network partitioning by using permissioned Blockchain.

## 8. BLOCKCHAIN TECHNOLOGY TO ENHANCE SECURITY IN MANETS

In [14],a framework is proposed for usingBlockchain technology to enhance security in MANETs. This approach modifies the existing Optimized Link State Routing Protocol (OLSR)protocol in MANETS to include blockchain and reputation. OLSR uses Multi point Relays (MPR) as relay nodes in routing. Here, the same nodes are used for sharing the blockchain with other MPR nodes. The blockchain managed by MPRs are called as MPR Blockchain. This is used to calculate the credibility of other nodes while selecting a route, thus avoiding malicious nodes.

## 9. FRAMEWORK FOR SUPPORTING CONNECTIVITY OF VANET/MANET

The paper [15] proposes a framework for supporting connectivity of VANET/MANET network nodes and elastic software configurable security services using blockchain with floating genesis block. The hindrance to use blockchains in MANET /VANET is the need to store the blockchain which consumes disk space due to its growth[14]. The authors of this article proposes a solution to this hindrance. There will be fixing blocks. Fixing blocks will store the details of the initial status of the system. So it can act as a genesis block for the next chain. This floating genesis blocks are digitally signed by trusted nodes to avoid any other cyber-attacks. This article also gives a comparison table about the existing approaches to solve the growth problem of Blockchain.

## 10. CONCLUSION

Blockchain models uses complex cryptographic mechanism for verification and trust management. But MANET devices are battery operated, and they cannot afford lot of computing intensive operations which are used in blockchain technology for securing operations. So when we use blockchain in MANET environment, there is a tradeoff between the security level, performance and the complexity. This paper discussed the existing approaches of applying Blockchain technology in MANETS especially in routing, trust management and scalability issues. The paper analyzed about the solutions given by various researchers and given review of using blockchain in MANETs. The future work will be to propose a framework for using blockchain in MANET environment by considering the tradeoffs.

## AUTHORS

Dr Sangheethaa S has completed her PhD from Anna University, India in Information and communication Engineering in the year 2012. She has completed ME in Network and Internet Engineering from Karunya University, India and BE in Information Technology from Bharathiyar University India. Her area of interest is Mobile ad hoc networks, security in mobile networks and recently her research area is in Blockchain. Currently she is working as Associate professor in the College of Information Technology, University of Fujairah, Fujairah, United Arab Emirates.

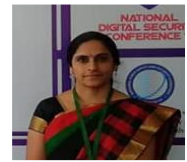